\documentclass[]{article}
\usepackage{amsmath}
\usepackage{amssymb}
\usepackage{amsfonts}
\usepackage{url}
\usepackage[affil-it]{authblk}
\usepackage{xfrac}
\newcommand{\ket}[1]{\left| #1 \right>} 
\newcommand{\bra}[1]{\left< #1 \right|} 

\begin{document}
\sloppy

\title{Simulating a perceptron on a quantum computer}

\author{Maria Schuld$^{a}$\footnote{Corresponding author: schuld@ukzn.ac.za }, Ilya Sinayskiy$^{a,b}$ and Francesco Petruccione$^{a,b}$}
\affil{$^{a}${\em{Quantum Research Group, School of Chemistry and Physics,  University of KwaZulu-Natal, Durban, KwaZulu-Natal, 4001, South Africa}}\\
$^{b}${\em{National Institute for Theoretical Physics (NITheP), KwaZulu-Natal, 4001, South Africa}}}\vspace{6pt}
\date{\today}

\maketitle

 Perceptrons are the basic computational unit of artificial neural networks, as they model the activation mechanism of an output neuron due to incoming signals from its neighbours. As linear classifiers, they play an important role in the foundations of machine learning. In the context of the emerging field of quantum machine learning, several attempts have been made to develop a corresponding unit using quantum information theory. Based on the quantum phase estimation algorithm, this paper introduces a quantum perceptron model imitating the step-activation function of a classical perceptron. This scheme requires resources in $\mathcal{O}(n)$ (where $n$ is the size of the input) and promises efficient applications for more complex structures such as trainable quantum neural networks.\\


\vspace{1cm}

\section{Introduction}
\begin{figure}[t]
  \centering    \includegraphics[width=0.25\textwidth]{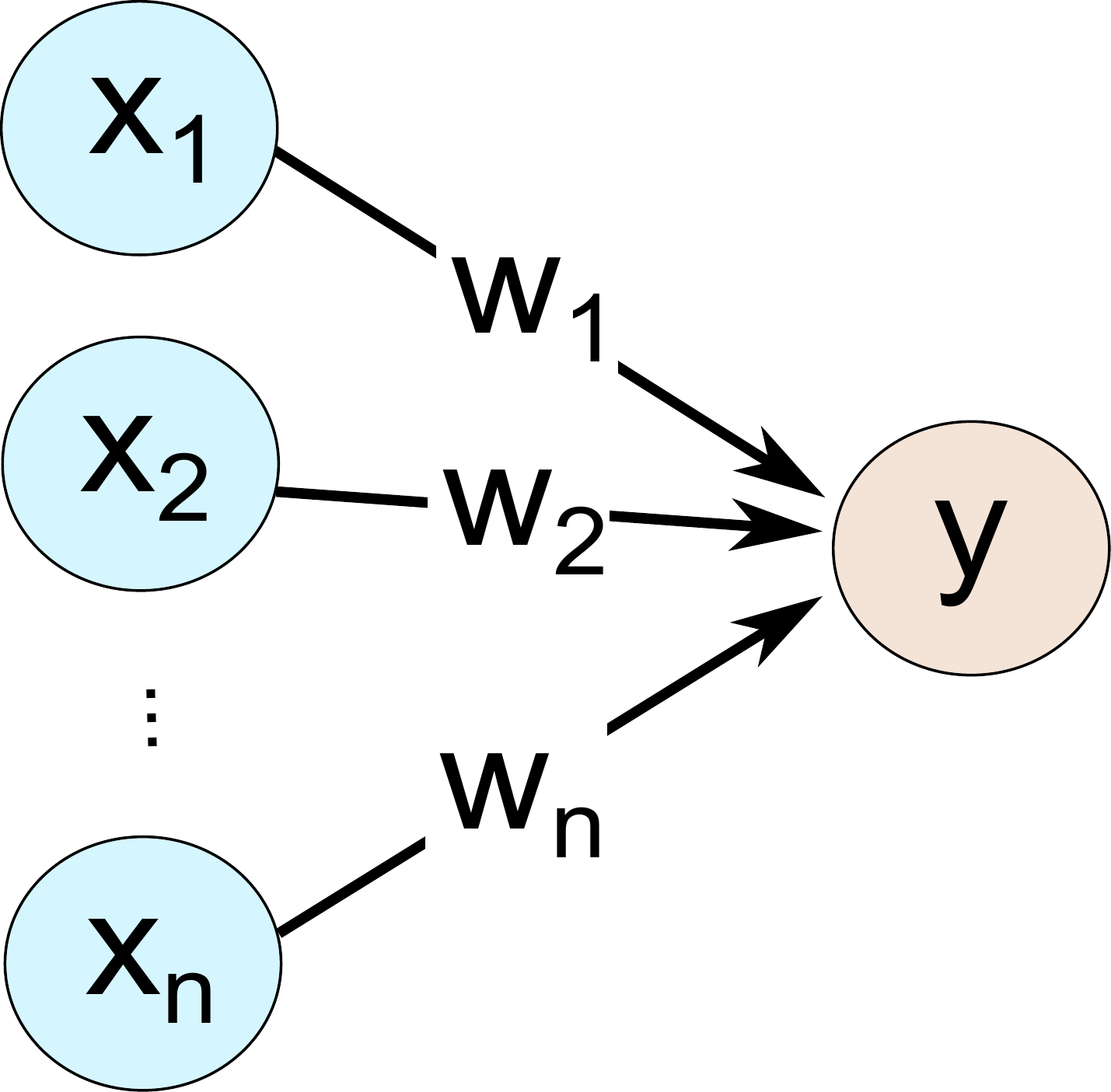} \hspace{2cm} \includegraphics[width=0.32\textwidth]{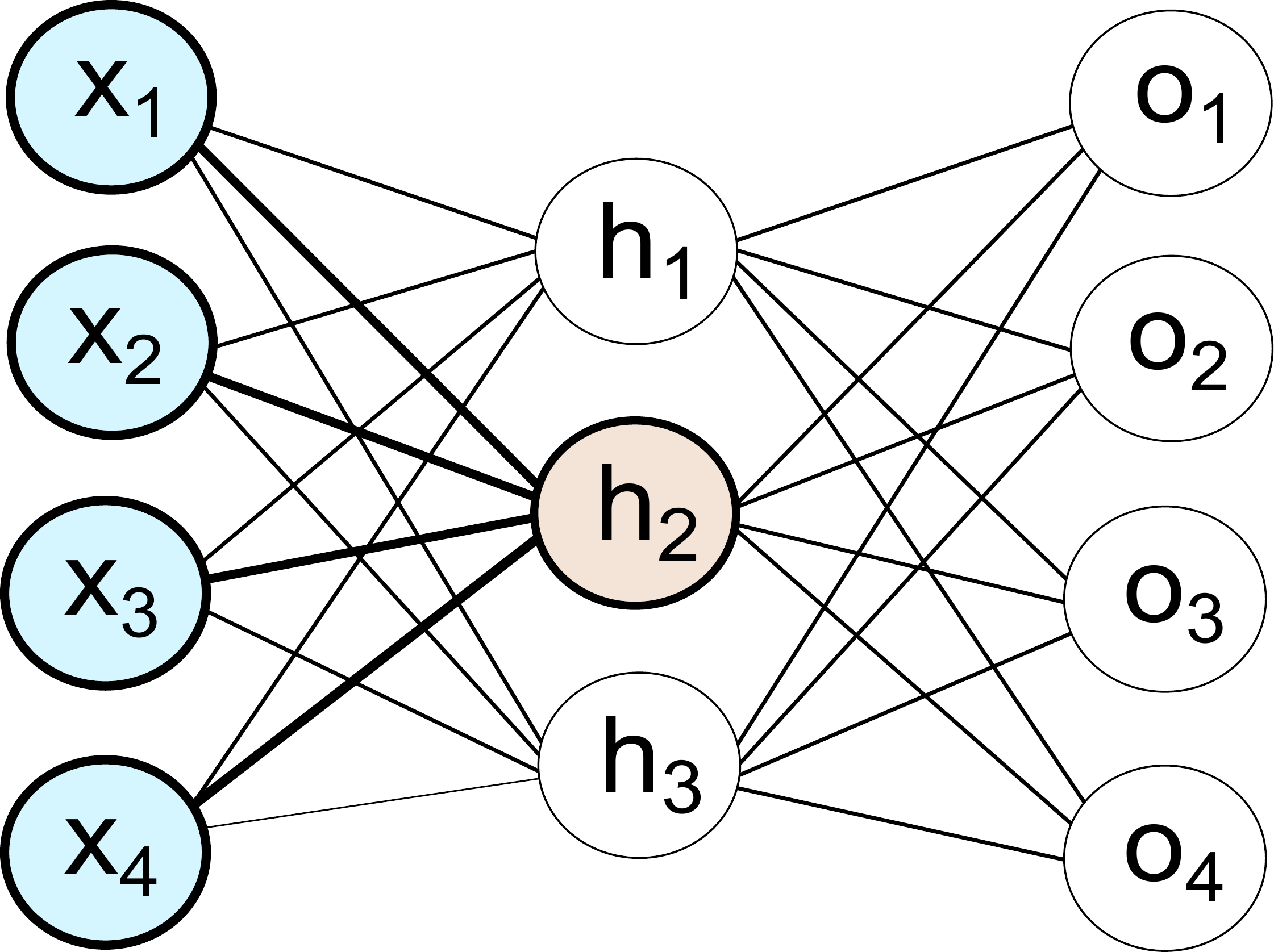} 
\caption{(Colour online) Left: Illustration of a perceptron model with input neurons $x_k = \{-1,1\}$, weights $w_k \in [-1,1)$,  $ k =1,...,n $  and output neuron $y \in  \{-1,1\}$. Right: Perceptrons are the basic unit of artificial neural networks (here a feed-forward neural network). The network has an input layer, one hidden layer and an output layer, which get updated in chronological order. Every node or neuron computes its value according to the perceptron activation function Eq (\ref{af}), so that the network maps an input ($x_1,...,x_4$) to an output ($o_1,...,o_4$). }
\label{figure1}
\end{figure}

\begin{figure}[t]
  \centering    \includegraphics[width=0.25\textwidth]{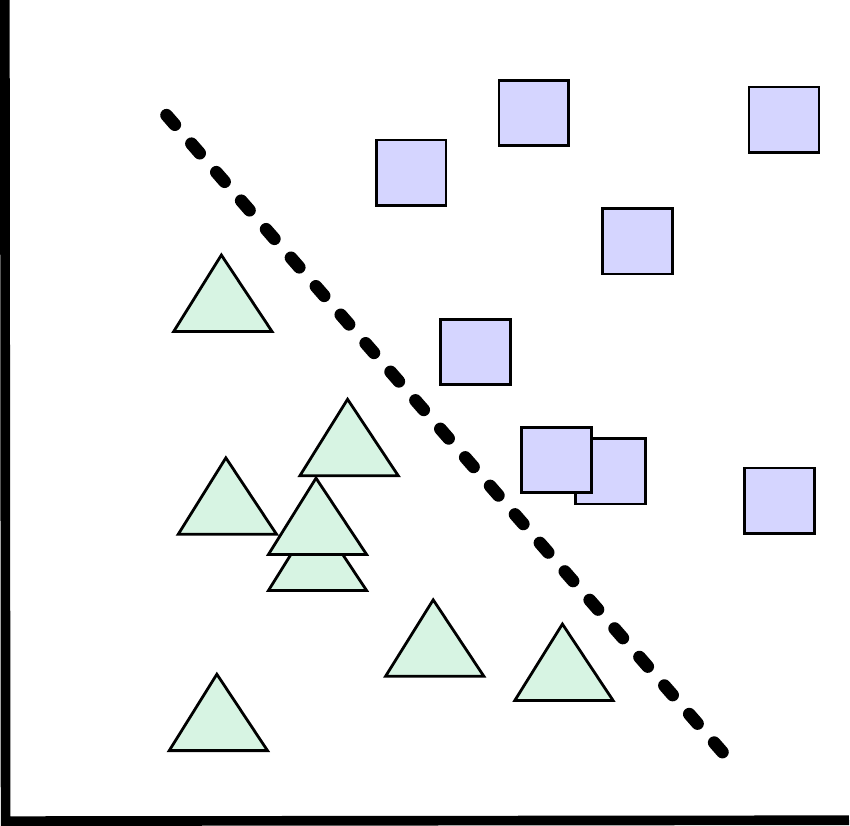} 
\caption{A dataset is linearly separable if it can be divided regarding its outputs by a hyperplane in phase space. }
\label{figure2}
\end{figure}

A perceptron is a mathematical model inspired by signal processing between neural cells that are assumed to be in either of the two states `active' or `resting'. It consists of $n$ input nodes called \textit{neurons} with values $x_k = \{-1,1\}, \; k =1,...,n $ that feed signals into a single output neuron $y$ (Figure \ref{figure1} left). Each input neuron is connected to the output neuron with a certain strength denoted by a weight parameter $w_k \in [-1,1)$ and the input-output relation is governed by the activation function 
\begin{equation} y =  \left\{   \begin{array}{l l}
		    1, & \quad \mathrm{if} \; \sum\limits_{k=1}^n w_{k} x_k \geq 0,\\
   		    -1, & \quad \mathrm{else.} 
  		\end{array} \right . . \label{af} \end{equation}
In other words, the net input $h(\vec{w},\vec{x})= \sum_{k=1}^n w_{k} x_k$ decides if the step-function activates the output neuron\footnote{Another frequent class of perceptrons use values $x_k = [-1,1], \; k =1,...,n $ and the logistic sigmoid activation function $y = \mathrm{sgm}(\sum\limits_{k=1}^m w_{k} x_k + \theta_y)$}. With their introduction by Rosenblatt in 1958 \cite{rosenblatt58}, perceptrons were a milestone in both the fields of neuroscience and artificial intelligence. Just like biological neural networks, perceptrons can learn an input-output function from examples by subsequently initialising $x_1,...,x_n$ with a number of example inputs, comparing the resulting outputs with the target outputs and adjusting the weights accordingly \cite{rojas96}. The high expectations of their potential for image classification tasks were disappointed when a study by Minsky and Papert in 1969 \cite{minsky69} revealed that perceptrons can only classify linearly separable functions, i.e. there has to be a hyperplane in phase space that divides the input vectors according to their respective output (Figure \ref{figure2}). An example for an important non-separable function is the $\mathrm{XOR}$ function. The combination of several layers of perceptrons to artificial neural networks (also called multi-layer perceptrons, see Figure \ref{figure1} right) later in the 1980s elegantly overcame this shortfall, and neural networks are up to today an exciting field of research with growing applications in the IT industry\footnote{Consider for example the latest developments in Google's image recognition algorthms \cite{lequoc13}.}.\\

Since two decades, quantum information theory \cite{nielsen10, plenio01} offers a fruitful extension to computer science by investigating how quantum systems and their specific laws of nature can be exploited in order to process information efficiently \cite{aharonov01, grover96}. Recent efforts investigate methods of artificial intelligence and machine learning from a quantum computational perspective, including the `quest for a quantum neural network' \cite{schuld14b}. Some approaches try to find a quantum equivalent for a perceptron, hoping to construct the building block for a more complex quantum neural network \cite{altaisky01,fei03,siomau14}. A relatively influential proposal to introduce a quantum perceptron is Altaisky's \cite{altaisky01} direct translation of Eq. (\ref{af}) into the formalism of quantum physics, namely $\ket{y} = \hat{F}\sum_{k=1}^m \hat{w}_{k} \ket{x_k}$, where the neurons $y,x_1,...,x_n$ are replaced by qubits $\ket{y},\ket{x_1},...,\ket{x_n}$ and the weights $w_k$ become unitary operators $\hat{w}_k$. The step activation function is replaced by another unitary operator $\hat{F}$. Unfortunately, this proposal has not been extended to a full neural network model. A significant challenge is for example the learning procedure, since the suggested rule inspired by classical learning, $\hat{w}_{k}^{[t+1]} = \hat{w}_{k}^{[t]} + \eta(\ket{d} - \ket{y^{[t]}}) \bra{x_k}$ with target output $\ket{d}$ and the learning rate $\eta \in [0,1]$, does not maintain the unitarity condition for the operators $\hat{w}_{k}$. Other authors who pick up Altaisky's idea do not provide a solution to this severe violation of quantum theory \cite{fei03, sagheer13, zhou07} (or propose an according open quantum systems framework, in which the operators still have to remain completely positivity and non-trace-increasing). Further models of quantum perceptrons can be found in the literature on quantum neural networks, but often remain vague in terms of the actual implementations \cite{gupta01, lewenstein94}, or do not apply quantum mechanics in a rigorous way \cite{kouda05, purushothaman97}. An interesting exception is Elizabeth Behrman's work introducing a perceptron as the time evolution of a single quantum object \cite{behrman00}, as well as Ricks and Ventura's ideas towards a superposition based learning procedure based on Grover's search algorithm \cite{ricks03}.\\

This contribution introduces a unitary quantum circuit that with only a small number of extra resources simulates the nonlinear input-output function of a classical perceptron as given in Eq. (\ref{af}). This quantum perceptron model has a high probability of reproducing the classical result upon measurement and can therefore be used as a classification device in quantum learning algorithms. The computational resources needed are comparable with the classical model, but the advantage lies in the fact that a quantum perceptron can process the entire learning set as a superposition, opening up new strategies for efficient learning. It can thus be seen as a building block of a more complex quantum neural network that harvests the advantages of quantum information processing. \\

\section{The quantum perceptron algorithm}
The quantum perceptron circuit is based on the idea of writing the normalised net input $\bar{h}(\vec{w},\vec{x})= \varphi \in [0,1)$ into the phase of a quantum state $\ket{x_1,...,x_n}$, and applying the phase estimation algorithm with a precision of $\tau$. This procedure will return a quantum state $\ket{J_1,...,J_{\tau}}$ which is the binary fraction representation of $\theta$ (or, equivalently, the binary integer representation of $j$ in $\theta = \frac{j}{2^{\tau}}$), which is in turn a good approximation for $\varphi$. More precisely, the output encodes the phase via $\theta = J_1 \frac{1}{2} + ... + J_{\tau} \frac{1}{2^{\tau}}$ (or $j = J_1 2^{\tau-1} + ... + J_{\tau} 2^0$) \cite{nielsen10}. The first digit of the output state of the quantum phase estimation algorithm, $J_1$, thus indicates if $\theta$ (and therefore with a good chance also $\varphi$) is bigger than $\frac{1}{2}$. The quantum perceptron consequently maps $(\vec{x}, \vec{w}) \rightarrow J_1$, which as we will see below reproduces the step activation function of a classical perceptron with a high probability.\\

\begin{figure}[t]
  \centering    \includegraphics[width=0.55\textwidth]{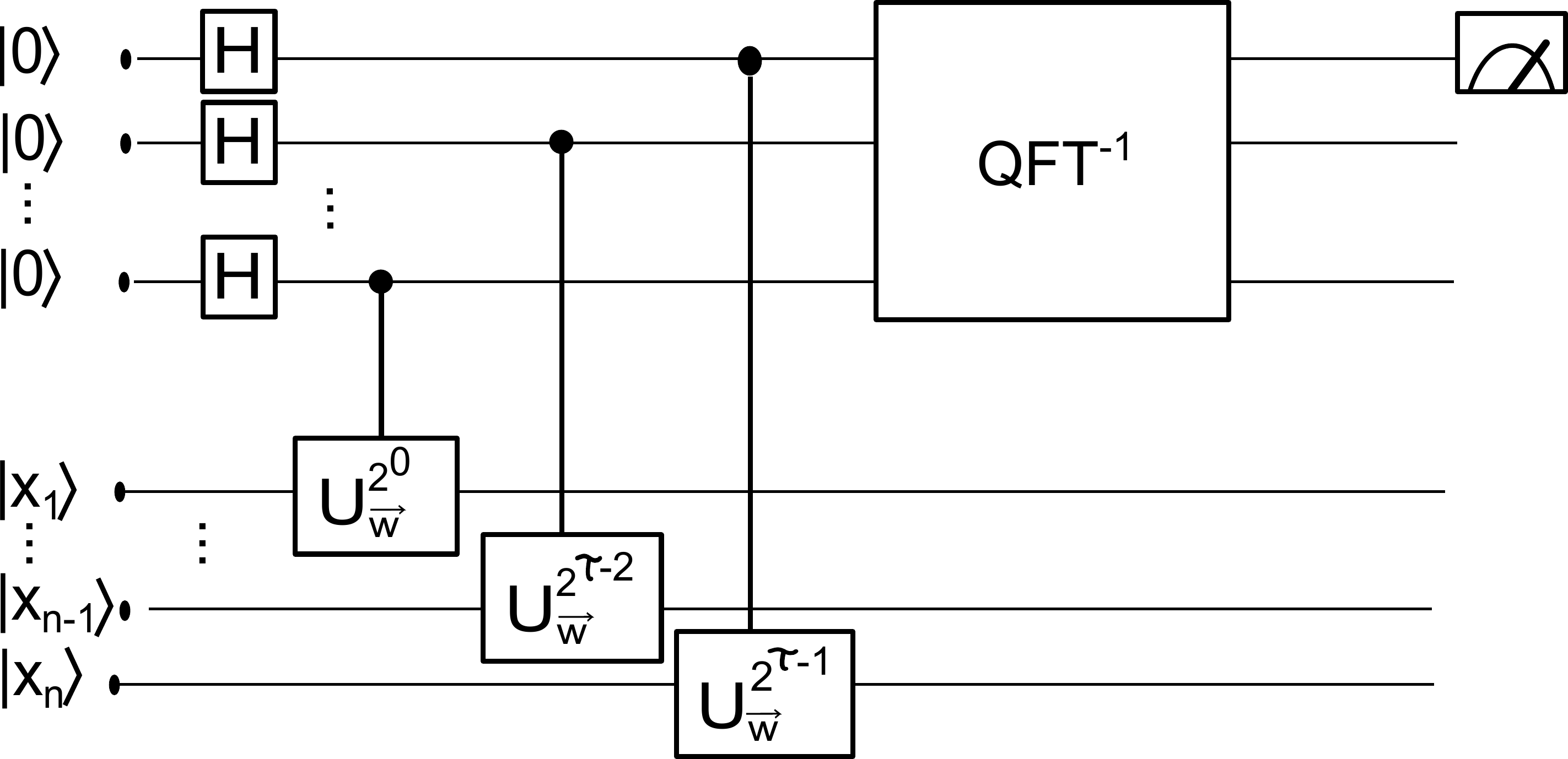} 
\caption{Quantum circuit for the quantum perceptron model. See also \cite{nielsen10}.}
\label{figure3}
\end{figure}

To give a more detailed impression of the quantum perceptron circuit (see also Figure \ref{figure3}), we assume an initial state $\ket{0,...,0}\ket{x_1,...,x_n} = \ket{0,...,0}\ket{\psi_0} $ composed of a register of $\tau$ qubits in state $0$ as well as an input register $\ket{\psi_0}$ with $n$ qubits encoding the binary states of the input neurons (note that in the quantum model, the $-1$ value is represented by a $0$ state). Hadamard transformations on the $\tau$ zeroes in the first register lead to the superposition $\frac{1}{\sqrt{2^{\tau}}} \sum_{j=0}^{2^{\tau}-1} \ket{J} \ket{x_1,...,x_n}$, in which $J$ is the binary representation of the integer $j$, and $\ket{J} = \ket{J_1,...,J_{\tau}}$.  We apply an oracle $\mathcal{O}$ that writes $j$ copies of a unitary transformation parameterised with the weights in front of the input register,
\begin{equation}\ket{J} \ket{\psi_0} \xrightarrow{\mathcal{O}} \ket{J} U(\vec{w})^j \ket{\psi_0}.  \label{step1}\end{equation}
The unitary $U$ writes the normalised input $\varphi$ into the phase of the quantum state. This can be done using the decomposition into single qubit operators $U(\vec{w}) = U_n(w_n)...U_2(w_2)U_1(w_1)U_0$ with each
\[U_k(w_k) = \begin{pmatrix}
e^{-2\pi i w_k \Delta\phi}&0\\
0&e^{2\pi i w_k \Delta\phi}
\end{pmatrix},\]
working on the input register's qubit $x_k$, and $\Delta{\phi} = \frac{1}{2n}$. $U_0$ adds a global phase of $\pi i$ so that the resulting phase of state $\ket{J} \ket{x_1,...,x_n}$ is given by $\exp(2\pi i (\Delta{\phi} h(\vec{w},\vec{x}) + 0.5) =  \exp(2\pi i \varphi)$. For learning algorithms it might be useful to work with parameters represented in an additional register of qubits instead of parametrised unitaries, and below we will give an according variation of the quantum perceptron algorithm. \\

The next step is to apply the inverse quantum Fourier transform \cite{watrous06, nielsen10}, $\mathrm{QFT}^{-1}$, resulting in
\begin{equation*} \frac{1}{\sqrt{2^{\tau}}} \sum \limits_{j=0}^{2^{\tau}-1} \exp^{2\pi i j \varphi} \ket{J} \ket{\psi_0}  \xrightarrow{\mathrm{QFT}^{-1}} \sum \limits_{j=0}^{2^{\tau}-1} \left(  \frac{1}{2^{\tau}} \sum \limits_{k=0}^{2^{\tau}-1} \exp^{2\pi i k  (\varphi - \frac{j}{2^{\tau}})}\right)  \ket{J}. \end{equation*}
In case the phase can be exactly expressed as $\varphi = \frac{j}{2^{\tau}}$ for an integer $j$, the amplitude of all states except from $\ket{J}$ is zero and the algorithm simply results in $\ket{J}$. For cases $\varphi \neq \frac{j}{2^{\tau}}$, it can be shown that in order to obtain $\varphi$ accurately up to $m$ bits of precision with a success probability of $1-\epsilon$, one has to choose $\tau = m + \lceil \log{(2+\frac{1}{2 \epsilon})}\rceil $ \cite{nielsen10}. Since we are only interested in the value of the first qubit, we would naively choose a precision of only $\tau = 2$ to obtain a $85\%$ probability of success. This would allow us to compute the quantum Fourier transform with minimal resources. \\

\begin{figure}[t]
  \centering    \includegraphics[width=0.54\textwidth]{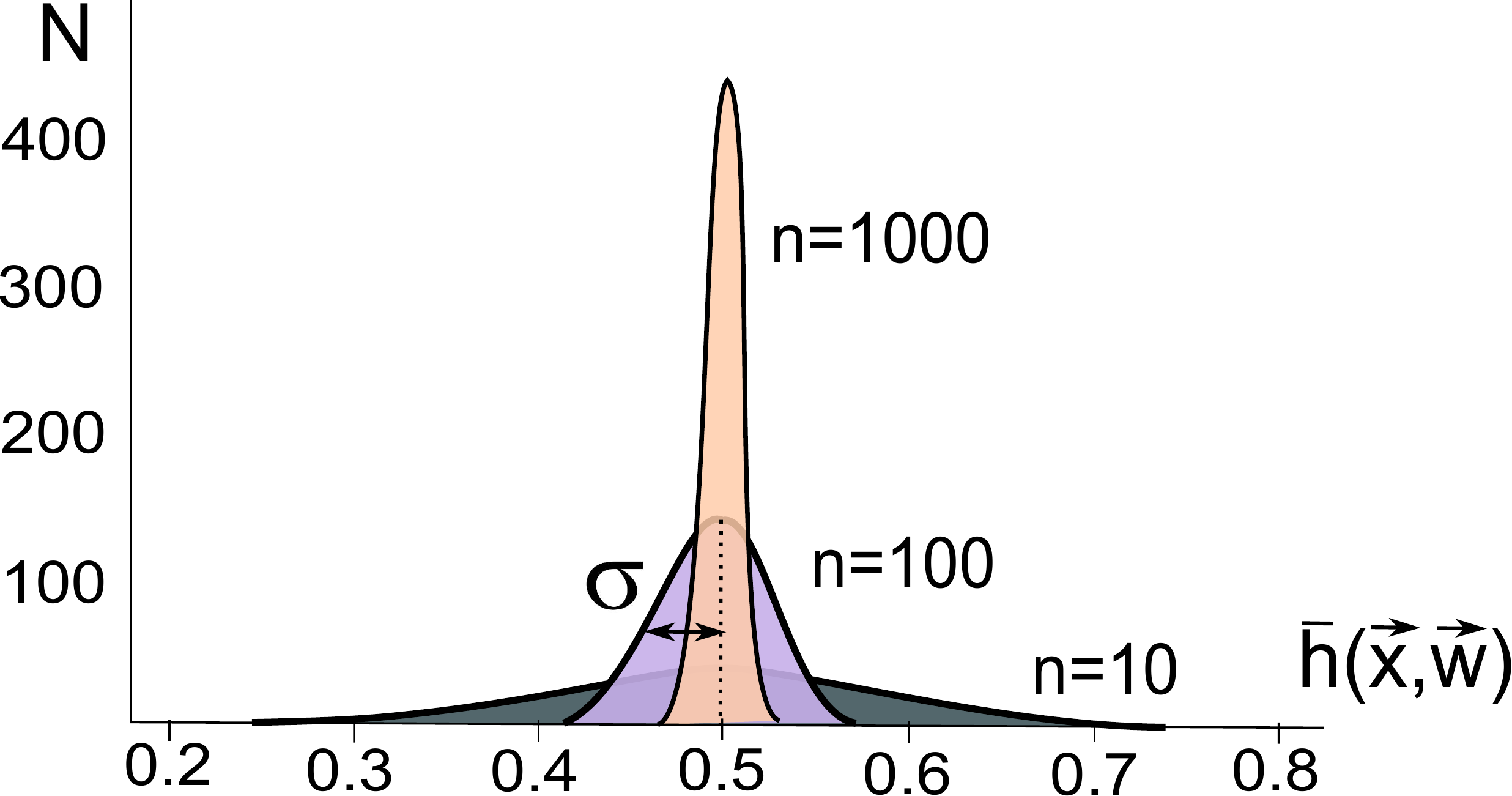} 
\caption{(Colour online) Histogram of distribution of values $\bar{h}(\vec{w},\vec{x})$ using random values for $\vec{w},\vec{x}$ with $10000$ data points. For $n=1000$ neurons, the distribution is much narrower in terms of the standard deviation $\sigma$ than for $n=10$. The precision of the algorithm consequently has to increase with the number of neurons.}
\label{figure4}
\end{figure}

However, it is important to note that the required size of $\tau$ needed can depend on the number of neurons $n$. To show this, let us assume a random distribution of binary values for the entries of $\vec{x}$ as well as random real values in $[-1,1)$ for $\vec{w}$. The higher the number of neurons $n$, the sharper the probability distribution of $\bar{h}(\vec{w},\vec{x})$ peaks around the average value of $\frac{1}{2}$ (Figure \ref{figure4}). This means that a perceptron unit has to have a higher resolution around this value, and we consequently need to increase the precision parameter $\tau$. Simulations show that for $n=10$ we need $\tau \geq 4$ to get a probability of more than $85\%$ to reproduce the classical perceptron's result,  while $n=100$ requires a precision of $\tau \geq 6$ and $n=1000$ a precision of $\tau \geq 8$. To quantify the relation between the number of binary digits $\tau$ and the number of neurons $n$, we assume (consistent with the simulations) that the standard deviation of the distribution of values for $\bar{h}(\vec{w},\vec{x})$ scales with $\sigma \sim  \frac{1}{\sqrt{n}}$. We require a precision $\tau$ that allows for a resolution in the order (e.g., a tenth) of the standard deviation, so that $\sigma \approx \frac{10}{2^{\tau}}$. The precision consequently scales as $\tau \sim \log{\sqrt{n}}$. Of course, these considerations are only true for random input variables and parameters, and we would expect a realistic case of a neural network to have its input values $\bar{h}(\vec{w},\vec{x})$  not necessarily distributed around $0.5$. But since the quantum perceptron might find application in the training of quantum neural networks, it is desirable that it can deal with almost random initial distributions over these values. It is therefore good news that the precision only grows logarithmically with the square number of neurons. \\

The computational complexity of the quantum perceptron algorithm is comparable to resources for the $n$ multiplications and single $\mathrm{IF}$-operation needed to implement a classical perceptron, which are in $\mathcal{O}(n)$. The quantum algorithm up to the inverse quantum Fourier transform requires $\tau + (n+1) \sum_k^{2^{\tau}-1} k$ elementary quantum gates\footnote{We can consider a set of elementary gates consisting of single qubit operations as well as the $\mathrm{CNOT}$ gate. \cite{barenco95}}. An efficient implementation of the  inverse quantum Fourier transform requires $\frac{\tau (\tau+1)}{2} + 3\frac{\tau}{2}$ gates \cite{nielsen10}. Taking $\tau$ as a fixed number we end up at a complexity of $\mathcal{O}(n)$. If we assume the above relationship between $\tau$ and $n$ derived from random sampling of $\vec{w}$ and $\vec{x}$, we still obtain $\mathcal{O}(n\; \mathrm{log}^2(\sqrt{n}))$, which is not a serious increase. A major advantage of the quantum perceptron is the fact that a quantum perceptron can process an arbitrary number of input vectors in quantum parallel if they are presented as a superposition $\sum_i \ket{x_i}$. The computation results in a superposition of outputs $\sum_i \ket{y_i}$ from which information can be extracted via quantum measurements, or which can be further processed, for example in superposition-based learning algorithms. The application of the quantum perceptron model will be discussed below. \\

As stated earlier, it can be useful to introduce a slight variation of the quantum perceptron algorithm, in which instead of parametrised operators, the weights $w_k, k=1,...,n$ are written into (and read out from) an extra quantum register. The initial state $\ket{\psi_0}$ in Eq. (\ref{step1}) thus becomes 
\[\ket{x_1,...,x_n; W^{(1)}_1,...,W^{(\delta)}_1 , \hdots , W^{(1)}_{n},...,W^{(\delta)}_n} = \ket{\vec{x}; \vec{w}}.\]
Consistent to above, $W^{(m)}_k$ is the $m$th digit of the binary fraction representation that expresses $w_k$ as $w_k = W^{(1)}_k \frac{1}{2} + ... + W^{(\delta)}_k \frac{1}{2^{\delta}}$ with a precision $\delta$. To write the normalised net input $\bar{h}(\vec{w},\vec{x})$ into the phase of quantum state $\ket{\vec{x}; \vec{w}}$ one has to replace the parameterised operator $U(\vec{w})$ in Eq. (\ref{step1}) with $\tilde{U} =  U_0 \prod_{k=1}^n \prod_{m=1}^{\delta}  U_{ W^{(m)}_k, x_k} $ where $U_0$ again adds $\sfrac{1}{2}$ to the phase and we introduce the controlled two-qubit operator
\[U_{W^{(m)}_k, x_k} = \begin{pmatrix}
1&0&0&0\\
0&1&0&0\\
0&0&e^{-2\pi i \Delta\phi \frac{1}{2^m}}&0\\
0&0&0&e^{2\pi i \Delta\phi \frac{1}{2^m}}
\end{pmatrix}.\]
The $m$th bit $W_k^{(m)}$ of the binary representation of $w_k$ controls the operation of shifting the phase by $-\Delta\phi \frac{1}{2^m}$ (for $x_k = 0$) or $\Delta\phi \frac{1}{2^m}$ (for $x_k = 1$), using $\Delta\phi$ from above.  Note that this implementation restricts the weights to $[0,1)$, but a sign for each parameter can be stored in an additional qubit, and its inverse $\mathrm{XOR}$ with $x_k$ can be used to control the sign of the phase shift.\\

\section{Application in quantum learning algorithms}

\begin{figure}[t]
  \centering    \includegraphics[width=0.64\textwidth]{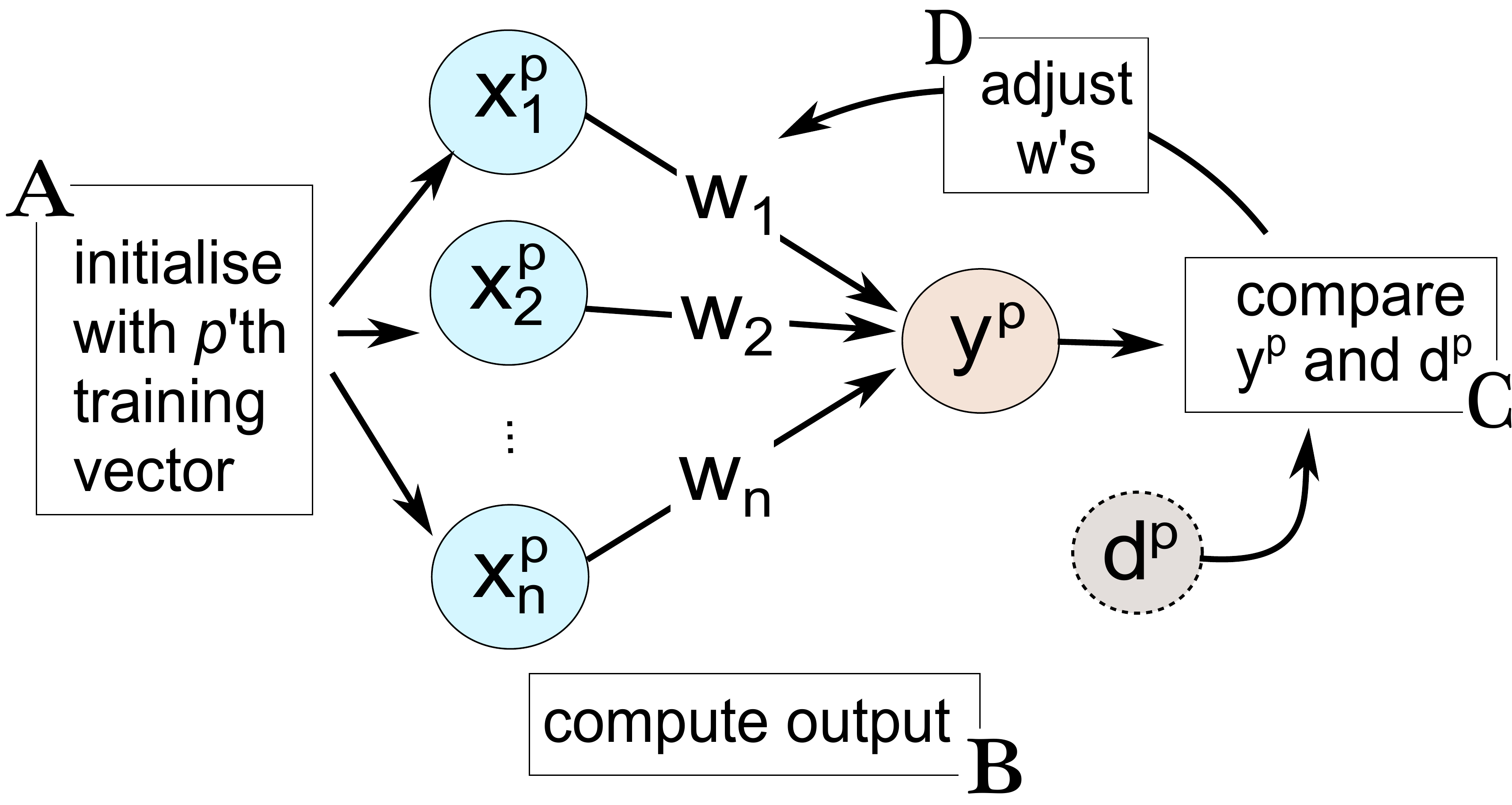} 
\caption{(Colour online) Illustration of one iteration in the classical perceptron training algorithm (the principle also holds for feed-forward neural networks constructed from perceptrons). A randomly selected training vector $\vec{x}^p = (x_1,...,x_n)^p$ from a training set is presented to the input layer (A) and the perceptron computes the actual output $y^p$ according to the perceptron activation function Eq (\ref{af}) using the current weights $w_1,...,w_n$ (B). The output  is compared with the given desired output $d^p$ for $\vec{x}^p$ (C) and the weights are adjusted to decrease the distance between the two (D). The quantum perceptron model can be applied to execute step B in the quantum versions of this training algorithm. }
\label{figure5}
\end{figure}

As mentioned before, perceptrons can be trained to compute a desired input-output relation by iteratively adjusting the weights when presented with training data (see Figure \ref{figure5}). The training data set $\mathcal{T} =  \{(\vec{x}^p,d^p)\}_{p=1,...,P}$ consists of examples of input vectors $\vec{x}^p$ and their respective desired output $d^p$. The actual output $y^p$ is calculated for a randomly selected vector $\vec{x}^p$ from this training set, using the current weight vector $\vec{w}$. The weights get adjusted according to the distance between $d^p$ and $y^p$,
\begin{equation}
\vec{w}'=\vec{w} + \eta (d^p-y^p)\vec{x}^p,
\end{equation}
where $\eta \in [0,1]$ is a given learning rate. By successively choosing random training examples, this procedure converges for linearly seperable problems to a weight vector that classifies all training examples correctly and can process new inputs as learned from the training set \cite{rojas96}.\\

While training a perceptron simply contains a number of classifications followed by vector addition, training a feed-forward neural network consisting of many interconnected perceptron units (see Figure \ref{figure1} right) quickly grows in terms of computational complexity since each output neuron indirectly depends on each weight of the previous layers. The most well-known training algorithm for feed-forward neural networks is based on gradient-descent \cite{rumelhart86} and changes the weights $w_{kl}$ between node $k$ and $l$ according to a very similar rule as for the percpetron, $w'_{kl} = w_{kl} - \eta \frac{\partial \mathrm{E}(\vec{o}^p - \vec{d}^p)}{\partial w_{kl}}$, where $\mathrm{E}$ is an error function depending on the computed output  $\vec{o}^p$ for a given input vector $\vec{x}^p$ of the training set and its target value $\vec{d}^p$. In other words, each weight is changed towards the steepest descent of an error function comparing the actual result with a target value. This procedure is called \textit{backpropagation} as it gets executed from the last to the first layer. There have recently been major improvements thanks to methods for efficient pre-training \cite{hinton06}, but the learning phase remains computationally costly for the dimensions of commonly applied neural networks.\\

A central goal of quantum neural network research is to improve the computing time of the training phase of artificial neural networks through a clever exploitation of quantum effects. Several training methods have been investigated, for example using a Grover search in order to find the optimal weight vector \cite{ricks03}, or using the classical perceptron training method to adjust a quantum perceptron's weight parameters\footnote{As mentioned in the introduction, an unresolved problem is to ensure that the operators remain unitary (or completely positive trace non-increasing).} \cite{altaisky01, zhou07}. Even though mature quantum learning algorithms are still a subject to ongoing research, from the examples it seems to be essential to generate an equivalent to the classical quality measure $\vec{d}^p-\vec{o}^p$ for the current weight vector $\vec{w}$. For this purpose a quantum perceptron unit is needed which maps input vectors $\ket{x^p}$ onto outputs $\ket{y^p}$ that can be compared with the target output. \\

The quantum perceptron is a model that  is able to calculate $\ket{y^p}$ equivalent to the classical model and with only very few resources. The difference to the classical model however is that it processes quantum information. This is not only a missing building block for the existing learning schemes mentioned above, but a basis on which to develop new quantum learning algorithms. For example, we currently investigate superposition-based learning algorithms, in which the training set is presented to the quantum perceptron as a superposition of feature vectors, and the quantum perceptron calculates the outputs in quantum parallel which can be further processed for learning. Such a scheme would be independent from the size of the training set.

\section{Conclusion}

The quantum perceptron model presented here offers a general procedure to simulate the step-function characteristic for a perceptron on a quantum computer, with an efficiency equivalent to the classical model. This fills a void in quantum neural network research, especially for quantum learning methods which rely on an equivalent to classical classification using quantum information. As a future outlook, the quantum perceptron model could be used to develop superposition-based learning schemes, in which a superposition of training vectors is processed in quantum parallel. This would be a valueable contribution to current explorations of quantum machine learning.\\

\section*{Acknowledgements}
This work is based upon research supported by the South African Research Chair Initiative of the Department of Science and Technology and National Research Foundation.


\end{document}